\documentstyle[11pt,epsfig]{article} 
\def\baselinestretch{1.3}
\newcommand{\ba}{\begin{array}}
\newcommand{\ea}{\end{array}}
\newcommand{\bd}{\begin{displaymath}}
\newcommand{\ed}{\end{displaymath}}
\newcommand{\be}{\begin{equation}}
\newcommand{\ee}{\end{equation}}
\newcommand{\bea}{\begin{eqnarray}}
\newcommand{\eea}{\end{eqnarray}}

\def\p{\pi}

\def\q2 {q^2}

\begin{document}
\begin{flushright}
\end{flushright}

\begin{center}
{\Large\bf  Bulk torsion fields in theories with large extra dimensions}\\[20mm]
Biswarup Mukhopadhyaya\footnote{E-mail: biswarup@mri.ernet.in},
Somasri Sen \footnote{E-mail: somasri@mri.ernet.in} \\
{\em Harish-Chandra Research Institute,\\
Chhatnag Road, Jhusi, Allahabad - 211 019, India} 

Soumitra SenGupta \footnote{E-mail: soumitra@juphys.ernet.in} \\
{\em Physics Department, Jadavpur University,\\
Kolkata - 700 032, India}\\[20mm] 
\end{center}

\vspace{0.5cm}
\vspace{0.5cm}

\begin{abstract}
We study the consequences of spacetime torsion coexisting with
gravity in the bulk in scenarios with large extra dimensions. Having
linked torsion with the Kalb-Ramond antisymmetric tensor field arising 
in string theories, we examine its artifacts on the visible 3-brane when 
the extra dimensions are compactified. It is found that while torsion 
would have led to parity violation in a 4-dimensional framework, all 
parity violating effects disappear on the visible brane when the torsion 
originates in the bulk. However, such a scenario is found to have 
characteristics of its own, some of which can be phenomenologically
significant.    
\end{abstract}

\vskip 1 true cm

\newpage
\setcounter{footnote}{0}

\def\baselinestretch{1.8}

\section{Introduction}

Although we apparently live in four-dimensional spacetime, 
some recent theoretical developments confront us with the tantalizing
possibility of extra spacelike compactified dimensions 
holding the key to the fundamental laws of physics. This, it is further 
argued, can happen  even at energy scales about as low as 
those probed by experiments so far. Among the variants of such theories,
two principal streams can be identified. These are the scenarios proposed
by Arkani-Hamed, Dimopoulos and Dvali (ADD) \cite{ADD} on one hand, and by 
Randall and Sundrum (RS) \cite{RS} on the other. Both of these approaches 
have sprung from a bid to solve the naturalness problem of the standard 
electroweak theory by recognizing scales arising out of compactified 
dimensions as the natural cut-offs on the standard model.

Some features are common to the two types of approach. The gravitational 
field tensor propagates in the higher dimensional manifold (called 
the `bulk') in each case. While the 
standard model (SM) fields are generally assumed to be confined to a 
3-dimensional `brane', some models with additional fields (for example, 
right-handed neutrinos)
in the bulk have also been proposed \cite{xxx}. In both cases \cite{ADD,RS} 
compactification of the 
extra spacelike dimensions creates a tower of massive fields on the 3-brane, 
out of erstwhile massless ones propagating in the bulk. The 
interaction of these
towers of states with the SM fields  provides the `new physics' inputs from
a phenomenological viewpoint. Such interactions, having 
manifold enhancement over the usual
gravitational effects via either the summation over a densely packed tower
of states (ADD) or a boost in the coupling by an 
exponential factor (RS), make such theories distinguishable in experiments.

In this paper, we examine the above scenarios in the context of
spacetime with torsion. First introduced into the fold of general relativity
in the Einstein-Cartan (EC) framework, torsion has come to be linked 
with matter fields having spin \cite{Saba}, just as curvature is 
connected with mass. 
A crucial difference brought about by torsion is the existence 
of an antisymmetric tensor part
in the affine connection \cite{Wald}. Some effects of torsion on 
theories with large extra
dimensions can be found, for example, in \cite{tatsu}. Here we take a 
more specific
approach where torsion is linked with the rank-2 antisymmetric tensor field 
$B_{\mu\nu}$, known as  
the Kalb-Ramond (KR) field, which occurs in the spectrum of heterotic 
string theories. The fact that most of the models under investigation 
are string-inspired lends legitimacy to such an approach and also makes it
possible to use certain properties of the KR field in making our predictions.

It has been demonstrated earlier \cite{ps,ssb} that torsion 
(occurring in the Lagrangian as an 
auxiliary field) gets equated to the KR field strength H on using the
equations of motion. Under such circumstances, the antisymmetric part in the 
affine connection is provided by the tensor $H_{\mu\nu\lambda}$, with

\begin{equation} 
 H_{\mu \nu \lambda} ~=~ \partial_{[\mu} B_{\nu \lambda]}
\end{equation}

Since both the symmetric tensor field of the graviton and the 
antisymmetric tensor field $B_{\mu\nu}$ control the geometry of spacetime,
we feel that it is natural to place them on identical footing by having
both of them exist in the bulk. In that case, the modification of the 
affine connection takes place in $(4+n)$ dimensions itself.

It has been shown, in the context of 4-dimensional
spacetime, that torsion destroys the cyclic property of the Riemann-Christoffel
tensor $R_{\mu\nu\alpha\beta}$, as a result of which a term of the form
$\epsilon^{\mu\nu\alpha\beta} R_{\mu\nu\alpha\beta}$ can be added to the 
scalar curvature \cite{hojman} in the 
gravitational action, causing the latter to be parity-violating. A consistent 
formalism was developed in \cite{bs} for incorporating the ensuing 
parity-violating effects in the interaction of different spin fields 
with spacetime curvature containing torsion. This was done by realizing that 
the modified connection in an EC theory can have a 
pseudo-tensorial part as well:

\begin{equation}
{\bar{\Gamma}_{\nu\lambda}}^{~\kappa}={\Gamma_{\nu\lambda}}^{\kappa}-
  {\frac{1}{M_P}}[{H_{\nu\lambda}}^{\kappa}-
q({\epsilon_{\nu\lambda}}^{\gamma\delta}
{H_{\gamma\delta}}^{\kappa}-{\epsilon_{\beta\lambda}}^{\kappa\alpha}
{H_{\nu\alpha}}^{\beta}+{\epsilon_{\beta\nu}}^{\kappa\alpha}
{H_{\lambda\alpha}}^{\beta})]
\end{equation}

\noindent
where $\Gamma_{\nu\lambda}^{\kappa}$ is the Christoffel symbol of 
Einstein gravity, symmetric in the two lower indices. The coefficient
${\frac{1}{M_P}}$ arises from dimensional requirements.   
Such a generalization of the covariant
derivative can be regarded as the most general way of incorporating
parity violation in the presence of torsion, since it not only yields the
added term $\epsilon^{\mu\nu\alpha\beta} R_{\mu\nu\alpha\beta}$ in
the scalar curvature but also leads to 
parity-violating effects for all matter fields with spin.
Here $q$ is a parameter determining the degree of parity violation.
Further implications of such parity violation has been explored in a number 
of recent works \cite{ssb,bs,sps,ssps,as}.

We now ask the question: what if the KR field coexists with gravity in the 
bulk, leading to a tower of antisymmetric tensor fields upon compactification?
More importantly, since parity-violation through the modified
connection may now  arise in higher dimensions, will its effects survive
in 4-dimensions as well? These and a few related questions are discussed
in what follows.  We shall outline our main arguments in the context of
an ADD scenario, although we shall also comment upon their validity for
RS-type theories. 

In section 2 we outline the basic features of such a theory 
in higher dimensions,
with special reference to a 6-dimensional framework.
The results of compactification of the extra dimensions and the 
new interactions obtained on the visible brane are discussed in section 3.
We summarise and conclude in section 4.

\section{Torsion in higher dimensional theories}

We begin by recalling the most important contentions 
of the ADD and RS types
of models. In ADD-type models \cite{ADD} , the compact and Lorentz degrees of
freedom can be factorized. The string scale $M_s$  (which can be 
as low as tens of TeV) controls the strength of gravity in $(4+n)$ dimensions,
and is related to the 4-dimensional Planck  scale $M_P$ by

\begin{equation}
\frac{R^n}{M_P^2} = (4\pi)^{n/2}\Gamma(n/2) M_s^{-(n+2)}
\end{equation}

\noindent
where  R is the compactification radius. The current limits on the departure  
from Newton's law of gravity at small distances are 
consistent with $R$ within a $mm$, for $n \ge 2$. Compactification
of the extra dimensions leads to a tower of Kaluza-Klein (KK) modes on the
brane where we reside. Thus a massless field in the bulk in general gives 
rise to a massive spectrum, the density of states being given by
\begin{equation}
\rho({m}_{\vec{n}})=\frac{R^n {{m}_{\vec{n}}^{(n-2)}}}{(4\pi)^{n/2}\Gamma(n/2)}
\end{equation}
where ${{m}_{\vec{n}}}={(\frac{4\pi^2{\vec{n}}^2}{R^2})}^{1/2}$ 
is the mass of a KK state with $\vec{n}=(n_1,n_2,........,n_n)$ \cite{han}. 

Consequently, in any process (involving the graviton, for example)
where a cumulative contribution from the
tower is possible, a summation over the tower of fields, convoluted
by the density, causes an enhancement, in spite of the suppression of
individual couplings by $M_P$. One thus expects appreciable contributions
to various processes at energies close to $M_s$. However, though it 
provides a stabilization of the electroweak scale, this type of a model
cannot avoid a still unexplained hierarchy between $1/R$ and $M_s$.

In the RS framework \cite{RS}, the last problem is ameliorated by introducing
a non-factorizable geometry.  The metric contains 
a `warp factor' which is an exponential function of
the compact dimension $\phi$: 

\begin{equation}
ds^2=e^{-2kr_c|\phi|}\eta_{\mu\nu}dx^{\mu}dx^{\nu}-r_c^2d\phi^2
\end{equation}

\noindent
where $r_c$ is the compactification radius on a $Z_2$ orbifold, and $k$ is
of the order of the higher dimensional Planck mass $M_5$. The standard model
fields are at $\phi~=~\pi$ whereas gravity propagating in the bulk peaks at
$\phi~=~0$. It turns out that  the 4-dimensional Planck mass $M_P$ in this
case is related to $M_5$ by

\begin{equation}
M_P^2=\frac{M_5^3}{k}[1-e^{-2kr_c\pi}]
\end{equation}

Furthermore, for $kr_c \simeq 12$, the exponential factor  generates
a mass scale of about a TeV from the Planck scale without requiring the 
postulate of an inordinately large compactification radius. The finite
renormalization of the tower  coming from any bulk field  
generates an additional factor of $e^{kr_c}$ in 
the coupling of any massive member of
the tower to the SM fields, although the tower itself remains rather sparse, 
having mass separations on the order of TeVs.

Let us now consider a scenario with torsion in bulk spacetime.
Following our earlier philosophy,
we wish to retain the possibility of parity violation in 
$(4+n)$ dimensions. Such a goal is attained via
the completely antisymmetric tensor density only if $n$ is even.
Let us further assume that torsion enters into
the geometry only through a `minimal coupling' scheme, being added 
linearly to the covariant derivative.

Our next observation is that a minimally coupled torsion can contribute
a pseudo-tensorial component to the affine connection
{\it only in 6-dimensions }({\it i.e.} for $n=2$), in the following way

\begin{equation}
{{\tilde{\Gamma}_{\nu\lambda}}}^{~\kappa}={\Gamma_{\nu\lambda}}^{\kappa}-
 {\frac{1}{M_s}}[{H_{\nu\lambda}}^\kappa - 
q{\epsilon_{\nu\lambda\alpha}}^{\kappa\rho\beta}{H_{\rho\beta}}^{\alpha}]
\end{equation}

The reason is obvious from the expression for the connection itself; the parity
violating (pseudo-tensorial) part must be of rank 3, and, with a rank 3 torsion
field strength available to us, the completely antisymmetric tensor 
density that one has to use here must be of rank 6. This constrains one to a
specific  dimensionality, namely, $n~=~2$. For $n > 2$, one has to introduce
terms in higher powers of $H_{\mu\nu\lambda}$ in the modified connection in
order to make it  parity violating.

With the modified affine connection defined in the above manner, it is
straightforward to calculate the scalar curvature in 6-dimensions:

\begin{equation}
R(g,H) =R(g)- {\frac{1}{M^2_s}}[H_{\mu\nu\lambda}H^{\mu\nu\lambda}-
2q \epsilon^{\mu\nu\lambda\alpha\beta\gamma}H_{\mu\nu\lambda}
H_{\alpha\beta\gamma}+q^2{\epsilon_{\nu\mu\lambda}}^{\omega\delta\rho}
{\epsilon_{\omega\delta\rho}}^{\alpha\beta\gamma}H_{\mu\nu\lambda}
H_{\alpha\beta\gamma}]
\end{equation}
\noindent
where the first term is the scalar curvature in 6-dimensions in the absence
of torsion. The second term is the extra piece arising in a Einstein-Cartan
picture. The third and fourth terms are the artifacts of the pseudo-tensorial
extension of the affine connection.

However, the relation  
$H_{\mu\nu\lambda} ~=~ \partial_{[\mu} B_{\nu \lambda]}$ immediately
implies that the third term in equation(8) is nothing but a surface term.
Thus we are led to  conclude that 
{\em $R(g,H)$ is invariant under parity} in spite of the pseudo-tensorial extension.
This can be attributed both to the way in which 
the KR field finds its way into the Lagrangian and to the restricted 
manner in which a rank-3 tensor can be combined with the rank-6 Levi Civita
tensor density to produce a modification to the connection.

Let us now specify the form of the 6-dimensional Lagrangian so as to
allow the interaction of the torsion field with matter fields on the visible
brane. If the $U(1)_{em}$  gauge field is to couple to torsion,
a consistent method is to extend  $H_{\mu\nu\lambda}$ by a Chern-Simons
term:

\begin{equation}
H_{\mu\nu\lambda} ~=~ \partial_{[\mu} B_{\nu \lambda]} 
~+~\frac{1}{M_s} A_{[\mu} 
F_{\nu\lambda]}
\end{equation}
\noindent
Here $F_{\nu\lambda}$ is the usual electromagnetic field tensor given
as $F_{\nu\lambda} = \partial_{[\nu}A_{\lambda]}$ and the corresponding
Lagrangian density for the electromagnetic field $A_{\lambda}$ is given in
the usual way in terms of this electromagnetic field tensor. Such a 
Chern-Simons term, invoked to achieve gauge anomaly cancellation in
heterotic String theory, produces a  gauge invariant interaction term
between torsion and the  gauge fields \cite{ps}. 
The implication of this term in 4-dimensional torsioned gravity has been 
examined in recent works on a number of issues, ranging from 
parity violation \cite{ps} to the rotation of the plane of polarization of
light \cite{sps,ssps}. 

With all the standard model fields confined to the 3-brane, 
the Chern-Simons term will contribute only when the 
indices attached to $H_{\mu\nu\lambda}$ 
correspond to the noncompact dimensions. There is a potential source of 
parity violation by virtue of the Chern-Simons term. The only addition in
$R(g,H)$ which can have this effect has to be linear in the completely
antisymmetric tensor density, and is of the form

\begin{equation}
\epsilon^{\mu\nu\lambda\alpha\beta\gamma}(\partial_{[\mu} B_{\nu \lambda]} 
A_{[\alpha} F_{\beta\gamma]} ~+~\partial_{[\alpha} B_{\beta \gamma]}
A_{[\mu} F_{\nu\lambda]}) 
\end{equation}
\noindent
This, however,  vanishes as a result of the antisymmetry of 
$\epsilon^{\mu\nu\lambda\alpha\beta\delta}$. Thus, unlike in the
4-dimensional case, the coupling of gauge fields to torsion
again turns out to be parity-conserving despite the pseudo-tensorial
part in the modified connection. However, there are additional
contributions to this interaction from the last term in
equation (8). We shall present these contributions in the next section 
where interactions in terms of the KK modes are listed.

Next, let us examine how a spin-1/2 field couples to torsion in
this kind of a scenario. For this we first need to write the free 
fermion Lagrangian in terms of the 6-dimensional Christoffel symbols
and the torsion tensor:

\begin{equation}
{\cal{L}}_D=\bar{\psi}[i\gamma^{\mu}(\partial_\mu-\frac{i}
{2}g_{\lambda\nu}\sigma^{ab}v^\nu_a\partial_\mu v^\lambda_b -
g_{\alpha\delta}\sigma^{ab}v^\beta_a v^\delta_b{\bar{\Gamma}_
{\mu\beta}}^\alpha)]\psi
\end{equation}
 
\noindent
where the $v^{\mu}_a$ are tetrads (here the Latin indices correspond 
to directions in the tangent space). It should be noted that the confinement
of fermions to the brane requires the indices answering to the
Dirac matrices to always correspond to the Lorentz ({\it i.e. } 
non-compact) dimensions. 
 
From above, one obtains

\begin{equation}
{\cal{L}}_D={\cal{L}}_E+{\frac{1}{M_s}}\bar{\psi}
[i\gamma^c \sigma^{ab}]\psi H_{cab}- 
{\frac{1}{M_s}}\bar{\psi}[iq\gamma^c\sigma^{ab}]
\psi{\epsilon_{cab}}^{\mu\nu\lambda} 
H_{\mu\nu\lambda}
\end{equation}

\noindent
As opposed to the cases with the 
scalar curvature and gauge field interaction, the fermion coupling to 
bulk torsion can thus violate parity as defined in 6-dimensions. 
This can be linked with the fact that the relevant 
terms involve a single power of the torsion field in this case.

To see whether the above Lagrangian entails parity violation in 
4-dimensions as well, one has to look at the KK towers of states on the 
3-brane. We investigate them in the next section.

\section{Interactions on the 3-brane}

In principle, compactification in an ADD scenario 
can give rise to a set of tensor fields
$\tilde{B}^{\vec{n}}_{\mu\nu}$, vector fields $B^{\vec{n}}_{\mu}$ and 
scalar fields $\chi^n$ in
4-dimensions. However, the bulk $B_{\mu\nu}$ can be assumed to be
block-diagonal in the compact and noncompact dimensions without any 
loss of generality. Besides, apart from sparing us the embarrassing
predicament of having massive vector fields in the low-energy spectrum,
this assumption is also consistent with the $SU(3)$ holonomy of the
Calabi-Yau manifold on which the process of compactification is
performed. Therefore, we shall consider only the tensor and scalar
fields on the visible brane. As regards
$\epsilon^{\mu\nu\lambda\alpha\beta\delta}$, two of its six indices
have to correspond to the compactified
dimensions, reducing it to its counterpart in 4-dimensions.

The first term in equation (8) reproduces Einstein gravity on the brane
together with the modifications caused by the tower of gravitons \cite{han}.
The remaining part of the modified scalar curvature in 6-dimensions 
yields the following kinetic and mass terms, 
corresponding to the tensor and scalar fields resulting from
$B_{\mu\nu}$, in the Larangian  ${\cal{L}}_{tor}$    :

\begin{eqnarray}
{\cal{L}}_{tor} =
&\sum_{\vec {n},\vec{n\prime}}&\{\partial_{[\mu}{\tilde{B}^{\vec n}}_
{~\nu\lambda]}\partial^{[\mu}{\tilde{B}}^{\vec{n}\nu\lambda]}-
6q^2({\rm det}~g^{\eta\eta\prime})\partial_{[\mu}{\tilde{B}^{\vec n}}_
{~\nu\lambda]}\partial_{[\alpha}{\tilde{B}^{\vec{n}}_{\beta\gamma]}}-
3(\frac{4\p^2\vec{n}^2}{R^2}){\tilde{B}^{\vec{n}}}_{\mu\nu}
{\tilde{B}^{\vec{n}\mu\nu}}\\\nonumber
&-&8q^2 (\frac{4\p^2\vec{n}.\vec{n\prime}}{R^2}){\tilde{B}^{\vec{n}}}_{\mu\nu}
{\tilde{B}^{\vec{n}\mu\nu}} + 6(1-q^2)
\partial_\lambda\chi\partial^\lambda\chi\}
\end{eqnarray}

\noindent
where $\eta(\eta\prime)$ runs over $\{\mu\nu\lambda(\alpha\beta\gamma)\}$.

Thus we have a tower of tensor fields, whose kinetic and mass terms can be
obatined in standard forms after proper rescaling and basis redefinition
(assuming a small $q^2$). 
However, an important point to note here is that the 
antisymmetry of $H_{\mu\nu\lambda}$ forbids any scalar mass term. One,
therefore, is left with {\it just one massless scalar $\chi$} in $(3+1)$ dimensions.
For $q^2~=~1$, the kinetic energy  term vanishes, leaving $\chi$ 
with no dynamical content.

As for the KK modes coupling to the $U(1)$ gauge field, both the second and
fourth terms in $R(g,H)$ become instrumental. After integrating out the 
compact dimensions, we thus obtain the following interactions:

\begin{equation}
{\cal{L}}_{tor-em} =
\frac{2}{M_p}\sum_{\vec n}[\partial_{[\mu}{\tilde{B}^
{\vec n}}_{~\nu\lambda]}A^{[\mu}F^{\nu\lambda]}-
6q^2({\rm det}~g^{\eta\eta\prime})\partial_{[\mu}{\tilde{B}^
{\vec n}}_{~\nu\lambda]}A_{[\alpha}F_{\beta\gamma]}]
\end{equation}
\noindent
where $\eta(\eta\prime)$ runs over $\{\mu\nu\lambda(\alpha\beta\gamma)\}$.
Therefore,  only the tensor tower couples with the gauge field 
via the Chern-Simons term.

The only place where the massless scalar $\chi$ exhibits some coupling 
in 4-dimensions is in the interactions with a fermion. On reduction
of the higher-dimensional Lagrangian given in equation (12), 
we have
\begin{equation}
{\cal{L}}_D={\cal{L}}_E+\sum_{\vec n} 
~i{\frac{1}{M_P}}\bar{\psi}\gamma^c \sigma^{ab}\psi~
\partial_{[c}{\tilde{B}^{\vec n}}_{~ab]}-
i {\frac{144qm}{M_P}}~\bar{\psi}\gamma_5~\psi~\chi
\end{equation}

\noindent
m being the mass of the fermion. Here the first term corresponds 
$\psi$ coupling to the KK tensor tower,
and the second, to the fermionic coupling of the massless scalar. 
The second term
arises purely due to the pseudo-tensorial extension in the
6-dimensional Lagrangian. The fact that the fermionic
current is confined to the 3-brane constrains all indices
of $H_{\mu\nu\lambda}$ to be Lorenzian if the latter is 
directly contracted with the fermionic current. However, from the
viewpoint of parity transformation in 4-dimensions, the above Lagrangian is
again invariant, as one can always use the phase freedom of 
the fields ${\tilde B}^{\vec{n}}_{\mu\nu}$ and  
$\chi$ independently on the 3-brane.

Thus we come to an important conclusion: {\it  even though the covariant 
derivative
in 6-dimensions can always be augmented with a pseudo-tensorial part
in presence of torsion, thereby causing parity violation in the bulk,
the ensuing theory in 4-dimensions turns out to be parity-conserving
in every sector.}

Several comments are in order here: 

\begin{itemize}
\item Although we have performed our analysis 
in the context of $(4+2)$ dimensions, the above conclusion  has wider 
applicability. The chain of arguments followed in this section and the 
previous one tells us that parity violation in 4-dimensions can arise 
from torsion field only when the KR field propagates in the 3-brane where
all the matter and gauge fields are confined. An exception to this is 
possible only when the affine connection is extended by a pseudo-tensorial
part consisting of higher powers of the KR field strength tensor.

\item In spite of the fact that parity violation disappears in 4-dimensions,
the pseudo-tensorial extension in equation (7) has non-trivial consequences
in this scenario. The fermionic coupling of the massless scalar field 
arises exclusively from this extension. Moreover, the pseudo-tensor
added in the bulk also modifies the gauge interaction of the tower
resulting from $B_{\mu\nu}$ on the visible brane. The possibility
of the massless scalar $\chi$ losing its dynamical content in the special
case of $q^2=1$ is also a consequence of parity 
violation in 6-dimensions.  

\item 
A string-inspired scenario also suggests the modification of 
the third rank antisymmetric field strength H by a gravitational 
Chern-Simons term in addition to the gauge Chern-Simons term \cite{gsw}:
\begin{equation}
H = dB - c_1\omega_Y - c_2\omega_L
\end{equation}
where $\omega_Y$ and $\omega_L$ are respectively the gauge and the 
gravitational Chern-Simons terms required for the gauge and gravitational 
anomaly cancellation and $c_1$, $c_2$  are constants. The gravitational 
Chern-Simons term however has not been  considered in the
foregoing discussion, since it is a higher 
derivative effect and in fact contains three derivatives where the 
others contain one. This term is therefore much more suppressed than the 
other terms and has been ignored in our analysis.

\item Although the above conclusions are derived in reference to an 
ADD-type model, they are mostly true in the 6-dimensional analogue of
an RS framework \cite{holdom} as well. This is because no 
recourse has been taken to any particular metric in the
reasoning leading to the disappearance of parity violation in the modified
scalar curvature and in gauge interaction of torsion fields. For 
fermion-torsion interactions, on the other hand, we have had to consider
the Lagrangian after compactification. There our arguments centrally depend
on (a) properties of the completely antisymmetric
tensor density, and (b) the fact
that the indices answering to the fermion current must be Lorenzian. 
All of the above points hold in an RS scenario, too, with 
appropriate effects coming from the warp
factor multiplied with the Minkowski part of the metric. 
A detailed investigation on an RS scenario in
this context will be reported later \cite{prog}.

\item  The presence of an additional weakly coupled massless (pseudo)scalar
may in general affect Big Bang Nucleosynthesis (BBN). An analysis of this 
problem has been performed in the second reference of \cite{ADD} in the
context of a torsion-free ADD model where, again, a massless scalar arises
from a bulk graviton as a result of compactification. There it has been
demonstrated that dissipation of the accumulated energy from the light
scalar(s) through Hubble expansion as well as freezing out of extra
dimensions well before the onset of the BBN era can be consistently 
accommodated. This, it has been further shown, is
possible with string scales of the order of $1-10~TeV$.Very similar
considerations apply to the light scalar in our case, ensuring that
it is safe from a BBN point of view. This is because the coupling of
both types of scalars (i.e. those arising out of graviton and torsion)
to matter is the same, being suppressed by the 4-dimensional Planck mass.
At the same time, one can avoid  overclosing the universe if the vacuum 
expectation value of each light scalar is small compared to the
inverse of its decay constant \cite{ekgada}.

\end{itemize}

We end this section by mentioning a few phenomenological consequences 
of a bulk torsion field. The first of these is the possibility of 
helicity flip of a fermion via scattering with the torsion field(s).
This possibility has already been studied in 4-dimensions where, however,
the coupling to torsion is suppressed by a factor of ${\frac{1}{M_P}}$
at each vertex \cite{as}.
With a KK tower of tensor fields ${\tilde B}^{\vec{n}}_{\mu\nu}$, the 
effective cross-section  
gets boosted upon integration over the entire tower of tensors.
Thus, using equation (15) the forward scattering cross-section for 
helicity flip of a fermion $f$ in the process
$f(p_1) {\tilde B}^{\vec{n}}(k_1) \longrightarrow f(p_2) {\tilde B}^{\vec{n}}(k_2)$ 
with $n=2$ is given by

\begin{equation}
\frac{d\sigma_{tot}}{{d\Omega}_{\theta=0}}=\frac{1}
{32\pi S}\int\int\frac{|\bar{p_2}|}{|\bar{p_1}|}|{\cal{M}}|^2 \rho
(m_{\chi}^2)\rho(m_{\chi\prime}^2)dm_{\chi}^2 dm_{\chi\prime}^2
\end{equation}
\noindent
where

\begin{eqnarray}
{\cal{M}}&=&\frac{1}{M_P^2}[\bar{u^+}(p_2)\gamma_\alpha
\gamma_5\frac{p_1\!\!\!\!\!/+
k_1\!\!\!\!\!/+m_f}{(p_1+k_1)^2-m_f^2}\gamma_\beta\gamma_5u^{-}(p_1)
(-k_{1\delta}k_{2\lambda})\epsilon^{\mu\nu\lambda\alpha}
\epsilon^{\rho\omega\delta\beta}\epsilon^{*}_{\mu\nu}(k_2)
\epsilon_{\rho\omega}(k_1)\nonumber\\
&+&\bar{u^+}(p_2)\gamma_\alpha\gamma_5\frac{p_1\!\!\!\!\!/-k_2\!\!\!\!\!/+
m_f}{(p_1-k_2)^2-m_f^2}\gamma_\beta\gamma_5u^{-}(p_1)(-k_{1\lambda}
k_{2\delta})\epsilon^{\mu\nu\lambda\alpha}\epsilon^{\rho\omega\delta\beta}
\epsilon_{\mu\nu}(k_1)\epsilon^{*}_{\rho\omega}(k_2)]
\end{eqnarray}

\noindent
is the amplitude for helicity flip. Here $k_1^2~=~m_\chi^2$
and $k_2^2~=~m_{\chi\prime}^2$, giving the masses of the initial
and final tensor states. The density of states $\rho$ can be expressed
in terms of $M_s$ and $M_P$ using equations (3) and (4), for $n = 2$.
The convolution with this density of KK states effectively replaces
the suppression factor $M_P$ by the string scale $M_s$ in the 
torsion-fermion coupling. Thus the cross-section picks up
an extra enhancement factor of
$({\frac{M_P}{M_s}})^4$ (leaving out the $m_{\chi,\chi\prime}$-dependence of 
the amplitude), so long as the integration over the tower is carried out
up to a mass scale on the order of $M_s$.

In a similar way, a boost can be expected in 
the forward scattering amplitude for flip from
negative to positive helicity for a neutrino 
when it is propagating against a background of a tower of KK torsion
states. This essentially means that an off-diagonal element
can arise in the Hamiltonian of a two-level system consisting of
an active and a sterile neutrino, thus resulting in active-sterile 
oscillation if  the neutrino has a Dirac mass. The signature of such a 
phenomenon can be 
in the form of a depletion in high-energy neutrino flux of cosmological
origin. Further details of the effect of a torsion tower on both neutrino
oscillation and the active-to-sterile scattering cross-section will
be presented in a subsequent paper.

\section{Summary and Conclusions}

We have performed a systematic analysis of the consequences of  
spacetime torsion in a scenario with large extra dimensions. Torsion
has been regarded here as arising from a massless Kalb-Ramond field
existing in the bulk. Working
in an ADD scenario, we have  included a pseudo-tensorial extension 
to the affine connection in the bulk, which is linear in the KR field 
strength. In a 4-dimensional framework, such an extension would have led 
to parity violation both in the modified scalar
curvature and in the coupling of torsion to matter fields with spin. 
However, when one starts from higher dimensions
and looks at the resulting 4-dimensional action involving
the KK towers of states, one gets back a parity-conserving theory.

The pseudo-tensorial part in higher dimensions is nonetheless found to be
of non-trivial consequence, since it gives rise to additional interaction
terms involving the KK tower of tensor states as well as the massless scalar 
obtained from the bulk KR field. We have also indicated that 
cumulative contribution from the tower can enhance 
helicity flip of a fermion when it is propagating in spacetime with
torsion. And finally, the conclusions drawn here are found to be by and
large applicable to an RS scenario as well, thus causing bulk
torsion to stand out as a distinctive phenomenon as far as low-energy
gravitational effects are concerned.

\noindent {\bf Acknowledgement:} We thank S. Naik and A. A. Sen for helpful
discussions. The works of BM and SSG have been partially supported by
grants from the Board of Research in Nuclear Sciences, Government of India.

\end{document}